\documentclass{ws-procs975x65}

\usepackage[footnotesize,bf]{caption}

\newcommand{\be}{\begin{equation}}
\newcommand{\ee}{\end{equation}}
\def\beq{\begin{eqnarray}}
\def\eeq{\end{eqnarray}}
\def\d{\mathrm{d}}

\begin{document}

\hspace{8.5cm} NIKHEF/2006-009

\title{HIGHER DIMENSIONAL VSI SPACETIMES AND SUPERGRAVITY}

\author{ANDREA FUSTER}

\address{NIKHEF, Kruislaan 409, 1098 SJ Amsterdam, The Netherlands \\
\email{fuster@nikhef.nl}}

\author{NICOS PELAVAS}

\address{Department of Mathematics and Statistics, Dalhousie University, Halifax, Nova Scotia B3H 3J5, Canada \\
\email{pelavas@mathstat.dal.ca}}

\begin{abstract}
We present the explicit form of higher dimensional VSI spacetimes
in arbitrary number of dimensions. We discuss briefly the VSI's
in the context of supergravity/strings.
\end{abstract}

\bodymatter

\section{Introduction}\label{intro}
We are interested in higher dimensional spacetimes for which curvature invariants of all orders vanish (VSI spacetimes), contained within this class are the higher dimensional pp-waves. In general, the higher dimensional VSI spacetimes have Ricci and Weyl type III \cite{Higher}. However, it is desirable to obtain explicit metric functions as has been done in four dimensions \cite{4DVSI}. We present metrics for Ricci type N, Weyl type III VSI spacetimes. 
\section{Higher dimensional VSI spacetimes}
Any $N$-dimensional VSI metric can be written in the form \cite{CMPPPZ,CSI}
\beq \d s^2=2\d u\left[\d
v+H(u,v,x^k)\;\d u+W_{i}(u,v,x^k)\;\d x^i\right]+\d x^i\d
x^i \label{Kundt}\eeq
where $u$, $v$ are light-cone coordinates and $x^i$, $i=1,\dots,N-2$, are real spatial coordinates. The functions $H$, $W_i$ are real-valued. Note that (\ref{Kundt}) is a subclass of the higher dimensional Kundt metrics \cite{CMPPPZ}. Restricting (\ref{Kundt}) to be of Ricci type N results in the Einstein equation $R_{uu}=\Phi$, where $\Phi$ is determined by the matter field (see Appendix). Two distinct cases arise depending on whether the functions $W_i$ depend on the light-cone coordinate $v$ ($\epsilon=1$) or not ($\epsilon=0$): 
\be
 W_1 = -\frac{2\epsilon}{x^1}\;v +W^{(0)}_1(u,x^k),\;\;\; 
 W_j = W^{(0)}_j(u,x^k) \nonumber
\ee
Here $j=2,\dots,N-2$; the superscript $(0)$ indicates functions without $v$-dependence. The case $\epsilon=W^{(0)}_i=0$ corresponds to higher dimensional pp-waves. It is perhaps surprising that only one of the $W_i$ functions is allowed to depend on $v$, as in the $4$-D case. The resulting spacetimes are summarized in Table $1$. An important remark is that the function $W^{(0)}_1$ can be gauged away by a coordinate transformation, the corresponding results have been recently presented \cite{vsipaper}. In this note we treat all $W^{(0)}_i$ functions equally, thereby revealing additional algebraic symmetries found in the metric functions of solutions. This complements the paper by Coley et al, wherein further details and explanation may be found.
\begin{table}[h] \label{table:1}
\begin{center}
\scriptsize{
\begin{tabular}{|l|c|c|c|} 
\hline
&&&\\
$\epsilon$ &  {\bf Weyl}  & {\bf Metric functions} & {\bf Eq.} \\
&&&\\
\hline
&&&\\

& III & $\begin{array}{c}   
W_{i}  =  W^{(0)}_{i}(u,x^{k})  \\ \\
H  =  H^{(0)}(u,x^{i})+\frac{1}{2}\left(F-W^{(0)i}_{\quad\quad,i}\right)v,\;\; F(u,x^i)\;\mbox{defined by}\; F,_i=\Delta W^{(0)}_{i} 
 \end{array}$ & \ref{app1} \\
&&&\\
\cline{2-4}
&&&\\
0 & N &  $ \begin{array}{c}
W_{i} =  x^{k}B_{ki}(u),\;\;B_{ki}(u)\;\mbox{antisymmetric}   \\ \\
H  =  H^{(0)}(u,x^{i}) \end{array}$ & \ref{app2} \\
&&&\\
\cline{2-4}
&&&\\
& O & $\begin{array}{c} W_i\;\mbox{as in type N} \\ \\ H^{(0)}=\frac{1}{2}W^iW_i+x^if_i(u)+H^{(0)}_{\Phi} \end{array}$ &  $\begin{array}{c} \ref{app2}  \\ \ref{app3} \end{array}$ \\
&&&\\
\hline
 &&&\\
& III & $\begin{array}{c} W_1  =  -\frac{2}{x^1}v  + W^{(0)}_{1}(u,x^{k}) \\ \\
W_{j}  =  W^{(0)}_{j}(u,x^{k}) \\ \\
H  =  H^{(0)}(u,x^{i})+\frac{1}{2}\left(\tilde{F}-W^{(0)i}_{\quad\quad,i}-\frac{2W^{(0)}_{1}}{x^1}\right)v+\frac{v^2}{2(x^1)^2},\\  \tilde{F}(u,x^i)\;\mbox{defined by} \; \tilde{F},_1=\frac{2}{x^1}W^{(0)i}_{\quad\quad,i}+\Delta W^{(0)}_{1},\; \tilde{F},_j=\Delta W^{(0)}_{j} \end{array}$ &  \ref{app4} \\
&&&\\
\cline{2-4}
&&&\\
1 & N &  $\begin{array}{c}  W_{1}  =  -2\frac{v}{x^{1}} + x^{j}B_{j1}(u)+C_{1}(u) \\ \\
W_{j}  =  x^{i}B_{ij}(u)+C_{j}(u),\;\; B_{j1}(u),B_{ij}(u)\; \mbox{antisymmetric} \\ \\
H  =  H^{(0)}(u,x^{i}) -\frac{W^{(0)}_{1}}{x^1}v+\frac{v^2}{2(x^{1})^{2}} \end{array}$  &  \ref{app5} \\
&&&\\
\cline{2-4}
&&&\\
& O &  $\begin{array}{c} W_i\;\mbox{as in type N} \\ \\ H^{(0)}=\frac{1}{2}(W^{(0)}_1)^2 +\frac{1}{2}\underset{j}{\sum} (W_j-x^1B_{1j})^2+x^1 g_0(u)+x^1x^ig_{i}(u)-\frac{1}{16}\Phi_0(u)x^1x^ix_i, \\ \Phi=\Phi_0(u)x^1 \end{array}$ &  \ref{app5} \\
&&&\\
\hline
\end{tabular}
}
\caption{All higher dimensional VSI spacetimes of Ricci type N. Ricci type O (vacuum) for $\Phi=0$ in (\ref{app1})-(\ref{app5}).}
\end{center}
\end{table}
\section{Outlook: VSI's in supergravity}
We can consider the embedding of VSI spacetimes in supergravity. The idea is to construct bosonic solutions of the supergravity equations of motion where the metric is that of a VSI\cite{gyrsugra,ortin,alanlett}. This has been extensively studied in the literature in the case of (Weyl type N) pp-waves. These are exact string solutions to all orders in $\alpha'$. The corresponding proof relies on the pp-waves invariants property. Furthermore, supergravity pp wave solutions preserve supersymmetry. The VSI metrics presented here generalize the pp-waves while the invariants property is maintained. Therefore it is natural to expect VSI supergravity solutions to have some of the pp-wave solutions' features. In fact, a certain class of VSI supergravity solutions has already been proved to be exact to all orders in $\alpha'$ \cite{alanlett}. Supersymmetry properties have only been studied in the case of VSI's with a covariantly constant null vector \cite{ortin}. In conclusion, these results provide an opportunity to explore VSI supergravity solutions, and their associated supersymmetries. \\

\appendix{Einstein equations}
{\small In the equations below $W_{ik}=W^{(0)}_{i,k}-W^{(0)}_{k,i}$ and $H^{(1)}$ is the coefficient of $v$ in the metric function $H$.
\begin{equation}
\;\;\;\;\;\;\;\;\;\triangle H^{(0)}- \frac{1}{4}W_{ik}W^{ik} -2  H^{(1)},_i W^{(0)i}-H^{(1)}W^{(0)i}_{\quad\quad,i}- W^{(0)i}_{\quad\quad,iu}+\Phi=0
\label{app1}
\end{equation}
\begin{minipage}[t]{0.49\textwidth}
\begin{equation}
\triangle H^{(0)}-2 \sum_{i<j} B_{ij}^2+\Phi=0 \label{app2}
\end{equation}
\end{minipage}
\hfill 
\begin{minipage}[t]{0.49\textwidth}
\be
H^0_{\Phi,_{ii}}=-\frac{1}{8}\Phi,\;\;\;H^0_{\Phi,_{ik}}=0 \label{app3} \ee 
\end{minipage}
\begin{eqnarray}
x^1 && \triangle \left( \frac{H^{(0)}}{x^1} \right) + \left( \frac{W^{(0)j}W^{(0)}_j-(W^{(0)}_1)^2}{x^1}\right),_{1} + \frac{2}{x^1}(W^{(0)}_1W^{(0)}_{1,1}-W^{(0)j}W^{(0)}_{1,j})  \nonumber \\ && -2  H^{(1)},_i W^{(0)i}-H^{(1)}W^{(0)i}_{\quad\quad,i} - \frac{1}{4}W_{ik}W^{ik}- W^{(0)i}_{\quad\quad,iu}+\Phi=0 \label{app4}
\end{eqnarray}
\be
x^1 \bigtriangleup \left( \frac{H^{(0)}}{x^1} \right)-\frac{1}{(x^1)^2}\left((W^{(0)}_1)^2+\underset{j}{\sum} (W_j-x^1B_{1j})^2\right) -\underset{j}{\sum} B_{1j}^2-2 \sum_{i<j} B_{ij}^2+\Phi=0  \label{app5}
\ee}

\vfill

\end{document}